\documentclass{ptapap}
\usepackage{floatrow}

\author{Matthew E. Shultz}[1]
\affil[1]{Annie Jump Cannon Fellow, Department of Physics and Astronomy, University of Delaware, 217 Sharp Lab, Newark, Delaware, 19716, USA}

\title{Modelling the Magnetic Fields and Magnetospheres of Early B-Type Stars}

\begin{document}

\maketitle

\begin{abstract}
The powerful radiative winds of hot stars with strong magnetic fields are magnetically confined into large, corotating magnetospheres, which exert important influences on stellar evolution via rotational spindown and mass-loss quenching. They are detectable via diagnostics across the electromagnetic spectrum. Since the fossil magnetic fields of early-type stars are stable over long timescales, and the ion source is internal and isotropic, hot star magnetospheres are also remarkably stable. This stability, the relative ease with which they can be studied at multiple wavelengths, and the growing population of such objects, makes them powerful laboratories for plasma astrophysics. The magnetospheres of the magnetic early B-type stars stand out for being detectable in every one of the available diagnostics. In this contribution I review the basic methods by which surface magnetic fields are constrained; the theoretical tools that have been developed in order to reveal the key physical processes governing hot star magnetospheres; and some important recent results and open-ended questions regarding the properties of surface magnetic fields and the behaviour of magnetospheric plasma. 
\end{abstract}

\section{Introduction}

The fraction of stars with a detectable surface magnetic field holds relatively steady at about 10\% from around spectral type A5 \citep{2019MNRAS.483.2300S} to the top of the Main Sequence (MS) \citep{2017MNRAS.465.2432G}. Hot stars lack a convective envelope, and there is therefore no mechanism by which a magnetic field might be sustained by a contemporaneous dynamo. For this reason, the magnetic fields of hot stars are believed to be magnetic `fossils', remnants of a previous period in the star's life \citep[e.g.][]{2015IAUS..305...61N}. The fossil field paradigm is bolstered by the properties of these magnetic fields: they are topologically simple \citep[typically dipolar, see e.g.][]{2019A&A...621A..47K}, strong \citep[a few hundred G to a tens of kG, e.g.][]{2007A&A...475.1053A,2019MNRAS.490..274S,2019MNRAS.483.3127S}, stable over a time-span of at least decades \citep[e.g.][]{2007A&A...475.1053A,2018MNRAS.475.5144S,2019MNRAS.483.3127S}, and exhibit no obvious correlations between e.g.\ rotational velocity and magnetic field strength \citep[e.g.][]{2007A&A...475.1053A,2019MNRAS.490..274S}. The stability of fossil fields over evolutionary timescales is backed up by magnetohydrodynamic (MHD) simulations \citep{2004Natur.431..819B}. 

Hot star atmospheres provide a ready ion source in the form of their powerful radiatively driven winds. When confined inside a strong magnetic field, the wind plasma animates a stellar magnetosphere \citep[e.g.][]{bm1997,ud2002}. This has direct consequences for the star's evolution, leading to rapid angular momentum loss as well as mass-loss quenching \citep[e.g.][]{wd1967,ud2002,ud2009}. Magnetospheres are furthermore detectable across the electromagnetic spectrum, from high-energy X-rays \citep[e.g.][]{oskinova2011} to low-frequency radio gyrosynchrotron \citep[e.g.][]{1987ApJ...322..902D,1992ApJ...393..341L}, yielding a wealth of observational diagnostics with which to probe different magnetospheric components. 


This contribution discusses the basic tools with which stellar surface magnetic fields are modeled (\S~\ref{sec:bfields}), and provides an overview of the theoretical frameworks within which various aspects of magnetospheric physics have been explored (\S~\ref{sec:magnetospheres}). It concludes with a summary of some of the most important recent results, and some of the key directions for future research (\S~\ref{sec:results}).

\section{Surface magnetic fields}\label{sec:bfields}

\subsection{The Oblique Rotator Model}

\begin{figure}
\floatbox[{\capbeside\thisfloatsetup{capbesideposition={right,top},capbesidewidth=5cm}}]{figure}[\FBwidth]
{\caption{Schematic of an oblique rotator model. The solid blue arrow indicates the rotation axis, inclined at an angle $i$ to the observor. The solid red arrow indicates the magnetic axis, tilted at an angle $\beta$ from the rotation axis. The strength of the surface magnetic field at the magnetic pole is $B_{\rm d}$.}\label{fig:orm}}
{\includegraphics[trim=100 0 100 400,width=5cm]{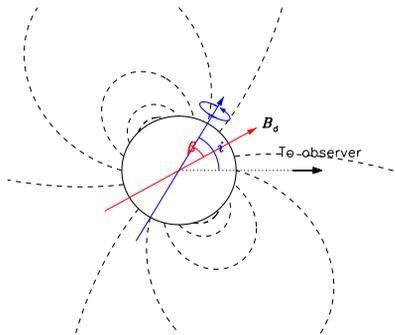}}
\end{figure}

\cite{1950MNRAS.110..395S} was the first to propose that the temporal variation of the disk-averaged longitudinal magnetic field $\langle B_{\rm z} \rangle$ could be most parsimoniously explained as a consequence of the rotation of a tilted dipole. In this {\em oblique rotator model} (ORM), illustrated in Fig.\ \ref{fig:orm}, the magnetic axis is tilted at an angle $\beta$ from the rotational axis, which is itself tilted at an angle $i$ with respect to the line of sight. As the star rotates, the angle between the line of sight and the magnetic axis changes, and $\langle B_{\rm z} \rangle$ varies sinusoidally. The variation in $\langle B_{\rm z} \rangle$ can then be reproduced if $i$, $\beta$, the strength of the magnetic dipole at the stellar surface $B_{\rm d}$, and the rotational period $P_{\rm rot}$ are known. 

There is a degeneracy between $i$ and $\beta$, which are related by the parameter $r = (\cos{\beta}\cos{i} - \sin{\beta}\sin{i}) / (\cos{\beta}\cos{i} + \sin{\beta}\sin{i})$ \citep{preston1967}. The $r$ parameter can also be determined from a harmonic fit to $\langle B_{\rm z} \rangle = B_0 + B_1\sin{(\phi + \Phi)}$, where $\phi$ is the rotation phase and $\Phi$ is a phase offset. Using this fit, $r$ can be equivalently written $r = (B_0 - B_1)/(B_0 + B_1)$. The usual means of determining $\beta$ is therefore to determine $B_0$ and $B_1$ from the $\langle B_{\rm z} \rangle$ curve, and then to constrain $i$ independently, typically via $P_{\rm rot}$, the stellar radius $R_*$, and the projected rotation velocity $v\sin{i}$. Once $i$ and $\beta$ are known, the surface strength of the magnetic dipole can then be obtained directly from these parameters, the limb darkening coefficient, and the maximum strength of $\langle B_{\rm z} \rangle$ \citep{preston1967}.

An ORM does not have to be purely dipolar, and in fact, in some cases a pure dipole provides a poor fit to the data. Early versions of non-dipolar ORMs consisted of either dipoles offset from the centre of the star, or in the linear superposition of quadrupolar and/or octupolar magnetic fields \citep[e.g.][]{1990ApJ...352L...5L,landmat2000}. While these methods were able to reproduce the $\langle B_{\rm z} \rangle$ variations of stars exhibiting significantly anharmonic curves, they generally provide a poor fit to the polarized line profiles, and have been superseded by inversion mapping. 

\subsection{Direct modelling of polarized line profiles}

With high resolution data, more sophisticated modelling of the Stokes $V$ profile can be performed. If the assumption of a purely dipolar magnetic field is maintained, a straightforward Bayesian approach developed by \cite{petit2012a} can place probabilistic constraints on the strength of the surface magnetic field regardless of the rotation phase, and can furthermore place robust upper limits on $B_{\rm d}$ when no magnetic field is detected and/or the period is unknown. 

The most sophisticated means of constraining the surface magnetic field topology and strength is via Zeeman Doppler Imaging \citep[ZDI;][]{pk2002}. This is a tomographic method that uses the variation in polarized line profiles across the rotational period to obtain a map of the star's magnetic field. An overview of this methodology is provided in these proceedings by Kochukhov. Here we note only that ZDI maps have generally confirmed that, with rare exceptions such as Landstreet's Star \citep{koch2011}, the magnetic fields of the majority of early-type stars are in fact mainly dipolar -- or more accurately, `twisted dipoles', i.e.\ dipoles with additional toroidal field components \citep{2019A&A...621A..47K}.

\section{Magnetospheres}\label{sec:magnetospheres}

\begin{figure}
\includegraphics[trim = 0 0 50 500, width=\textwidth]{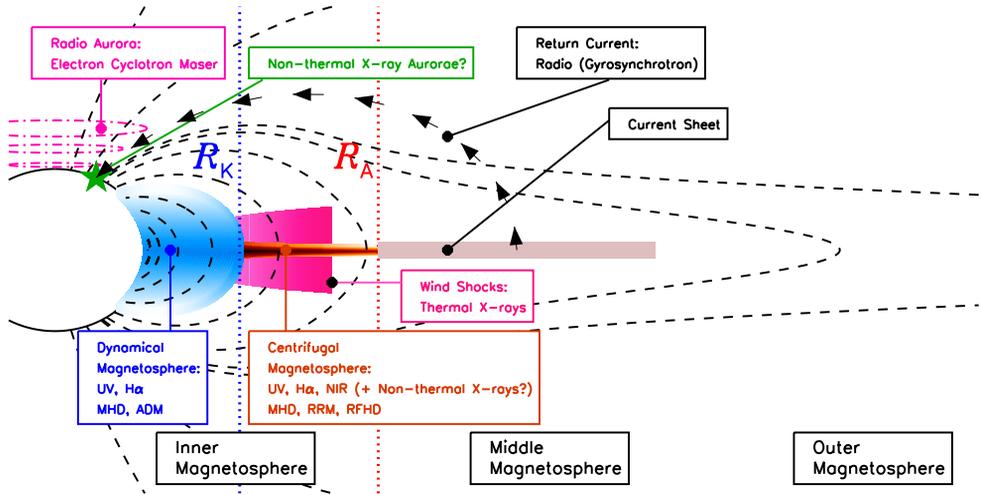}
\caption{Schematic of a stellar magnetosphere for an aligned dipole around a rapidly rotating star. Labels indicate the name for each component, the diagnostics by which it is detectable, and (where relevant) the theoretical formalisms employed in modelling.}
\label{fig:magnetosphere}
\end{figure}

Hot stars have powerful line-driven winds with mass-loss rates ranging from $10^{-10}$ to $10^{-5}~{\rm M_\odot~yr^{-1}}$ and terminal velocities of 1000 to 2000 ${\rm km~s^{-1}}$. Inside a magnetic field, the ionized wind plasma can be channeled to flow along magnetic field lines, thus providing an ion source for the formation of a magnetosphere. A schematic illustration of a stellar magnetosphere is provided in Fig.\ \ref{fig:magnetosphere}. Whether or not a magnetosphere will form depends on the relative strengths of the wind and the magnetic field, quantified via the wind magnetic confinement parameter $\eta_*$, which is essentially the ratio of the kinetic energy density in the wind to the magnetic energy density, evaluated at the magnetic equator at the stellar surface \citep{ud2002}. If $\eta_* > 1$ the wind is magnetically confined. Since a dipolar magnetic field declines with distance $r$ from the star as $1/r^3$, while the wind kinetic energy increases with distance due to line driving, magnetic confinement inevitably fails past a certain point. The radius at which the wind opens the magnetic field lines is the Alfv\'en radius $R_{\rm A}$, which can be determined via a straightforward scaling with $\eta_*$ \citep{ud2008}. 

The collision of wind flows from opposite magnetic colatitudes at the tops of magnetic loops leads to the formation of magnetically confined wind shocks \citep[MCWS;][see pink shaded region in Fig.\ \ref{fig:magnetosphere}]{bm1997,ud2014}. These wind shocks in turn generate thermal X-ray emission, with the result that magnetic hot stars are more X-ray-luminous by about 1 dex as compared to non-magnetic hot stars \citep[e.g.][]{2014ApJS..215...10N}. 

An important consequence of magnetospheres is that they act as extended moment arms, rapidly removing angular momentum from their host stars \citep[e.g.][]{wd1967,ud2009}. The rate of angular momentum loss increases with $R_{\rm A}$ and with the mass-loss rate. 

The region inside the closed field lines is referred to as the inner magnetosphere (see Fig. \ref{fig:magnetosphere}). Since the inner magnetosphere is dominated by the magnetic field, the Lorentz force enforces corotation of the magnetospheric plasma with the magnetic field; because the magnetic field is effectively fixed, the plasma corotates with the photosphere, out to a distance of up to tens of stellar radii. 

\cite{petit2013} established a basic division of hot magnetic stars into two classes based on the structure of the inner magnetosphere. The first class possesses only a {\em dynamical magnetosphere} (DM; blue shaded region in Fig.\ \ref{fig:magnetosphere}). In a DM, the rotation of the star is negligible and the motion of the wind plasma is dominated by line driving and gravity. Radiative acceleration pushes plasma to the tops of magnetic field loops, and plasma is then pulled back to the stellar surface by gravity. Therefore, a DM can only be detectable if the stellar wind is able to fill it faster than it empties on dynamical timescales. In the UV, DMs are detectable for essentially all magnetic OB stars (since UV lines are a $\rho$ diagnostic), whereas in H$\alpha$ (a $\rho^2$ diagnostic) DMs are generally only detectable around the magnetic O-type stars\footnote{With the notable exception of the magnetic B0\,III star $\xi^1$ CMa \citep{2017MNRAS.471.2286S}.}. The strong winds of O-type stars mean that they spin down extremely fast, such that they almost exclusively possess DMs\footnote{So far the only exception to this is Plaskett's Star \citep{2013MNRAS.428.1686G}}. 

Corotation can lead to significant centrifugal forces acting within the inner magnetosphere. At the Kepler corotation radius $R_{\rm K}$ gravitational and centrifugal forces are equivalent \citep[e.g.][]{town2005c,ud2008}. If $R_{\rm K} < R_{\rm A}$ then there exists a region in the inner magnetosphere within which gravitational infall is prevented by centrifugal support. This gives rise to the second class described by \cite{petit2013}, stars that also have a {\em centrifugal magnetosphere} (CM; orange shaded region in Fig.\ \ref{fig:magnetosphere}). Within the CM plasma is able to build up to much higher density than is possible within the DM. As a result of this, numerous B-type stars possess CMs detectable in H$\alpha$ despite their relatively weak winds. CM emission is also detectable in the NIR \citep[e.g.][]{2014ApJ...784L..30E,2015A&A...578A.112O}. 

Outside $R_{\rm A}$ corotation breaks down and the wind opens the magnetic field lines, leading to the formation of a current sheet powered by the stellar wind (grey shaded region in Fig.\ \ref{fig:magnetosphere}). This is the {\em middle magnetosphere}. The middle magnetosphere's current sheet accelerates electrons to relativisitic velocities. Some of these return to the star along the magnetic field lines, as illustrated in Fig.\ \ref{fig:magnetosphere}. Upon re-entering the inner magnetosphere these electrons emit incoherent gyrosynchrotron radio emission \citep{2004A&A...418..593T}. Some of the electrons can become trapped within auroral circuits (pink ovals in Fig.\ \ref{fig:magnetosphere}), leading to strongly circularly polarized, beamed radio emission emitted nearly perpendicularly to the magnetic axis \citep[][see also Das et al., this volume.]{2000A&A...362..281T}, a similar phenomenon to that seen in pulsars. 

\begin{table}
\caption[]{Summary of available diagnostics for various classes of magnetic early-type stars. }
\label{obstab}
\begin{tabular}{l | c c c c}
\hline\hline
Diagnostic & A5--A0 & B9--B6 & B5--B0 & O9--O5 \\
\hline
X-ray & No & Yes & Yes & Yes \\
UV & No & No & Yes & Yes \\
H$\alpha$ & No & No & Yes & Yes \\
NIR & No & No & Yes & Yes \\
Radio & No & Yes & Yes & No \\
\hline\hline
\end{tabular}
\end{table}

Table \ref{obstab} summarizes the observed magnetospheric characteristics of various sub-populations of hot stars with fossil magnetic fields. The magnetic A-type stars are not known to display any magnetospheric activity, consistent with such stars having negligible winds that do not provide a sufficient ion source to populate the circumstellar region. Late B-type stars do not show either UV, H$\alpha$, or NIR, but do show both X-ray and radio. This suggests that the weak winds of this sub-population are unable to fill their magnetospheres sufficiently to become detectable at intermediate (UV to NIR) wavelengths, but are strong enough to produce thermal X-rays and gyrosynchrotron. The magnetic O-type stars are detectable using all diagnostics save radio, for the simple reason that their radio photospheres are much larger than their Alfv\'en radii and effectively swallow any gyrosynchrotron radiation that may be emitted within their inner magnetospheres \citep{2015MNRAS.452.1245C}\footnote{In fact, a flat radio spectrum was reported for Plaskett's Star by \cite{2017MNRAS.465.2160K}, but as this system is a binary it cannot yet be ruled out that this might be due to a colliding wind shock.}.

The only class of stars for which the full suite of diagnostics are detectable is the early B-type stars. Since each diagnostic probes a different region of the magnetosphere, for these stars it is uniquely possible to reveal their magnetospheres in full detail. 

\subsection{The dynamical magnetosphere}





About 20\% of the early B-type stars are too slowly rotating to possess a CM \citep{2019MNRAS.490..274S}. However, rapidly rotating stars still possess DMs in the innermost regions of their inner magnetospheres. The DMs of early B-type stars are often detectable via UV emission \citep[e.g.][]{petit2013}, and are furthermore capable of generating thermal X-rays via MCWS. 

The fundamental dynamics of DMs have been explored via both 2D and 3D MHD simulations. \cite{ud2002} first investiated the effects of magnetic wind confinement on mass-loss quenching using 2D simulations, and revealed the snake-like patterns of infall characterizing plasma transport within this region. 3D simulations of aligned rotators by \cite{ud2013} showed that this turbulent infall leads to stochastic line strength variations consistent with those observed in the H$\alpha$ variations of the magnetic O-type star $\theta^1$ Ori C. The effects of dipole tilt in a star with moderate rotation ($R_{\rm K} > R_{\rm A}$) were examined in 3D simulations by \cite{2019MNRAS.489.3251D}, who found that dipole tilt introduces a rotational modulation of the thermal radio emission and, hence, can affect mass-loss rates inferred via this diagnostic.

Numerical limitations place strong restrictions on the range of magnetic confinement strengths that can be simulated using MHD, with the maximum for 2D models currently being around $\eta_* = 1000$, whereas the magnetospheres of magnetic B-type stars can easily reach $\eta_* \sim 10^6$ \citep[][]{2019MNRAS.490..274S}. Even within the range of $\eta_*$ values for which simulations are practical, they are time-consuming, making it difficult to apply them to large populations with a range of fundamental and ORM parameters. In consequence, analytic treatments have been developed which build upon the results from MHD simulations. 

The first of these was the X-ray Analytic Dynamical Magnetosphere (XADM) model developed by \cite{ud2014}, which determines the locations of MCWS based upon the strength and speed of the wind and the strength of the confining magnetic field. Comparison with observed X-ray luminosites revealed a generally good agreement across several dex \citep{2014ApJS..215...10N}, although the observed luminosities are typically about 10\% of their predicted values; this is believed to be a consequence of self-absorption of the X-rays by the magnetospheric plasma. 

A more general Analytic Dynamical Magnetosphere (ADM) model was developed by \cite{2016MNRAS.462.3830O}. This model incorporates the free-flowing wind above the magnetic poles, as well as upflows and downflows along magnetic field lines, in order to determine the spatial, density, and velocity distributions of the circumstellar plasma. This enables the ADM model to predict photometric variability due to occultation of the star by the DM \citep{2019MNRAS.tmp.2573M}, and when coupled with radiative transfer, to predict line emission at arbitrary wavelengths \citep{2019MNRAS.483.2814D}. 

\subsection{The centrifugal magnetosphere}

A majority of magnetic early B-type stars have CMs, and about 1/4 have CMs detectable in H$\alpha$ \citep{2019MNRAS.490..274S}. 

As with DMs, CMs have been explored using both MHD simulations and analytic models. So far only 2D MHD simulations have been published (however, see ud-Doula et al., this volume). \cite{ud2006,ud2008} verified the formation of a compressed, high-density disk above the Kepler corotation radius, and explored the phenomenon of `centrifugal breakout', whereby mass-loading of the CM by the wind eventually causes the confining magnetic field lines to rupture, leading to explosive ejection of plasma away from the star. Angular momentum loss within rotating magnetospheres was examined by \cite{ud2009}, results which led to the development of an angular momentum loss scaling law depending on mass-loss rate, rotational velocity, and magnetic confinement strength. 

The magnetic confinement strengths and rotational velocities of the majority of early B-type star CMs are beyond the numerical ability of MHD simulations to explore, making analytic treatments of high importance. The most influential of these is the Rigidly Rotating Magnetosphere (RRM) model developed by \cite{town2005c}. RRM models calculate the gravitocentrifugal potential along each magnetic field line, and map out an accumulation surface by locating the potential minima. Plasma is assumed to collect in hydrostatic equilibrium on the accumulation surface. For a tilted dipole, the accumulation surface is a warped disk about halfway between the magnetic and rotational equators, with the densest pockets of plasma being located at the intersections of these two equators. Since there is no rotational support below $R_{\rm K}$, the region between the star and $R_{\rm K}$ is empty. Corotation of the plasma furthermore leads to a linear increase in rotational velocity with distance from the star. The RRM model therefore predicts an emission signature with no emission below $R_{\rm K}$, and two emission bumps at $R_{\rm K}$ (typically, 2 or 3 $R_*$ or equivalently, 2 or 3 times $v\sin{i}$). Stellar rotation then leads to a characteristic `double helix' pattern of emission variability. 

This basic pattern of emission has been verified for several magnetic early B-type stars \citep[e.g.][]{leone2010,bohl2011,oks2012,grun2012,rivi2013,2015MNRAS.451.1928S,2015ApJ...811L..26W,2016MNRAS.460.1811S,2016ASPC..506..305S,2017MNRAS.465.2517W,2017A&A...597L...6C}, and is so recognizable that it has enabled the discovery of new magnetic stars on the basis of their emission signatures alone \citep{2014ApJ...784L..30E}.

If the assumption of a dipolar magnetic field is relaxed, an arbitrary RRM (aRRM) model can be calculated via potential field extrapolation from e.g.\ a magnetic map obtained via ZDI. This has so far been done only for $\sigma$ Ori E. Using this method \cite{2015MNRAS.451.2015O} were able to simulatenously reproduce the star's H$\alpha$ emission profile variability and the photometric eclipses of the star by the dense CM plasma. However, their synthetic light curve was unable to reproduce the out-of-eclipse variation, despite including the photometric variability predicted by its surface chemical abundance patches. They speculated that its CM plasma might not only be dense enough to eclipse the star \citep[as explored by][]{town2008}, but also to scatter light when projected next to the star \citep[see also][this volume]{2019arXiv191204121M}.

The assumption of hydrodynamic equilibrium along each field line is relaxed in the Rigid-Field Hydrodynamic (RFHD) simulations developed by \cite{town2007}, which effectively treats each magnetic field line as a rigid pipe along which the wind plasma can freely flow. RFHD provides predictions not just for the H$\alpha$ and photometric signatures of a CM, but also for X-ray and ultraviolet variability. \cite{2014ApJS..215...10N} found that some of the rapidly rotating magnetic B-type stars are about 1 dex overluminous in X-rays as compared to XADM predictions, a discrepancy that might be explained by the additional X-ray luminosity caused by the release of centrifugal potential energy \citep{town2007}. Tilted dipoles in RFHD were explored by \cite{2016MNRAS.462.3672B}, who found that magnetic topologies and rotation can strongly affect the properties of the stellar wind along each field line. 

Mass balancing via breakout as investigated by \cite{ud2006,ud2008} is the subject of some controversy. The X-ray flares that might be expected to accompany breakout events have not been detected. No sign of large-scale re-organization of the CM has been seen, and RRM modeling of magnetospheric eclipses by \cite{town2013} yielded a lower limit on the CM mass about 2 dex below the breakout mass derived by \cite{town2005c}. This discrepancy has led to the suggestion by \cite{2018MNRAS.474.3090O} that mass balancing might instead be achieved via a steady-state leakage mechanism involving diffusion and drift across magnetic field lines.

\section{Recent Results}\label{sec:results}

\subsection{Magnetic fields and stellar evolution}

Fossil magnetic fields are in principle stable on evolutionary timescales \citep[e.g.]{2004Natur.431..819B,2010ApJ...724L..34D}, however it has long been questioned whether this is truly the case, or whether the magnetic flux $\Phi = B_{\rm d}R_*^2$ might decline over time. The first serious attempt to investigate this was conducted by \cite{2006AA...450..763K}, who examined cool Ap/Bp stars and concluded that $\Phi$ was essentially constant. \cite{bagn2006} and \cite{land2007,land2008} looked at Ap/Bp stars in clusters, which have well-defined main-sequence turnoff ages, and found evidence that $\Phi$ may decline for the more massive stars in their sample. \cite{2016A&A...592A..84F} compared the occurrence of magnetic vs.\ non-magnetic stars across the upper MS, and found that the incidence of magnetic stars declined rapidly with age for the more massive stars in the sample, inferring from this that $\Phi$ might decline more rapidly for more massive stars. However, direct comparison of the magnetic upper limits for non-detected O-type stars to the field strengths of the detected O-type stars by \cite{2019MNRAS.489.5669P} showed that the existing spectropolarimetric survey data is not sufficiently precise to rule out flux conservation at the top of the MS.

These studies were limited in that ORM parameters were unavailable for the majority of their samples. Two recent studies have examined the question of flux conservation using samples of magnetic hot stars with well-characterized fundamental and ORM parameters. \cite{2019MNRAS.483.2300S,2019MNRAS.483.3127S} conducted a volume-limited study of all Ap stars below about 3 M$_\odot$ within the nearest 200 pc, and found that $\Phi$ is apparently constant with this sample, consistent with the results of \cite{land2007}. \cite{2018MNRAS.475.5144S,2019MNRAS.485.1508S,2019MNRAS.490..274S} looked at the population of early B-type stars, between about 5 and 20 M$_\odot$, and found that $\Phi$ decreased for all stars, and decreased more rapidly for the most massive stars, consistent with both the results of \cite{land2007} for the more massive end of their sample and the conclusion of \cite{2016A&A...592A..84F}. 

\cite{2019MNRAS.490..274S} also examined rotational evolution, finding that rotation periods increase dramatically over time. This result is qualitatively conistent with magnetic braking, but direct comparison of stellar and spindown ages showed that the former are, for much of the sample, up to 2 dex longer than the latter. Their simultaneous derivation of magnetospheric parameters for the sample \citep[accomplished using a self-consistent Monte Carlo Hertzsprung-Russell diagram sampler incorporating all available atmospheric, magnetic, rotational, and other constraints, described in detail in Appendix A of][]{2019MNRAS.490..274S} showed that H$\alpha$ emission from CMs requires both very strong magnetic confinement and very rapid rotation. Since magnetic fields weaken and rotation slows over time, the H$\alpha$-bright stars are also all young; indeed, they constitute a 2/3 majority in the first third of the MS. 

Notwithstanding the strong evidence that magnetic stars experience rapid magnetospheric braking over evolutionary timescales, efforts to directly measure angular momentum loss have yielded mixed results. Only one star, $\sigma$ Ori E, shows unambiguous spindown consistent with magnetic braking timescales \citep[][Petit et al., in prep.]{town2010}. CU\,Vir and Landstreet's Star instead show alternating spin-up and spin-down \citep[][]{miku2008,miku2011,2017ASPC..510..220M,2019arXiv191204121M}, while HD\,142990 was recently shown to be rapidly spinning up \citep{2019MNRAS.486.5558S}. No definitive explanation for this behaviour has yet been determined, although suggestions include torsional oscillations \citep{2017MNRAS.464..933K} or internal differential rotation \citep{2018CoSka..48..203M}.

The evolutionary models used for all of these studies were calculated for non-magnetic stars, and therefore do not account for the effects of mass-loss quenching \citep[potentially very signficant for O-type stars, e.g.][]{2017MNRAS.466.1052P}, angular momentum loss, and likely internal effects of magnetic fields such as inhibition of mixing and enforcement of solid-body rotation. The first generation of models to self-consistently incorporate all of these effects using a variety of 1D stellar evolution codes were developed by \cite{2019MNRAS.485.5843K} and Keszthelyi (submitted), who found generally reasonable agreement between the observed and predicted rotational properties of the population of magnetic B-type stars. 

\subsection{Magnetospheres}

The RRM model provides an excellent description of the H$\alpha$ variations of single magnetic stars. The first tidally locked binary with H$\alpha$ emission consistent with an origin in a CM, HD\,156324, was reported by \cite{2018MNRAS.475..839S}. This star's  H$\alpha$ emission shows no evidence of a second bump, which \citeauthor{2018MNRAS.475..839S} suggested may be due to modification of the gravitocentrifugal potential by the introduction of a close binary companion, raising the question of whether a relatively straightforward modification of the RRM formalism might be able to reproduce the emission properties of this star.

A strongly variable X-ray light curve has been reported by \cite{2017A&A...602A..92P} for $\rho$ Oph A, a rapidly rotating magnetic \citep{2018A&A...610L...3P} B-type star. This is a striking discovery, because the X-ray light curves of most magnetic stars are either not variable, or not strongly variable \citep{2014ApJS..215...10N}, which is believed to be a consequence of the production of X-rays in optically thin plasma several stellar radii from the photosphere \citep[e.g.][]{town2007,ud2014}. 

Aside from stochastic variations in the magnetospheric diagnostics of magnetic O-type stars, hot star magnetospheric variations are entirely dominated by rotational modulation. A surprising exception to this is the magnetic $\beta$ Cep pulsator $\xi^1$ CMa, which shows {\em pulsational} modulation of both its X-ray light curve \citep{2014NatCo...5E4024O} and its H$\alpha$ profile \citep{2017MNRAS.471.2286S}. The former is probably related to the variation in the star's mass-loss rate throughout its pulsation cycle, however the latter appears to also be related to redistribution of plasma within the star's DM. 

Gyrosynchrotron emission has received increasing attention in the past few years. \cite{2015MNRAS.452.1245C} and \cite{2017MNRAS.465.2160K} observed a large sample of magnetic OB stars, and while they detected several of the B-type stars, they did not detect a convincing signature of gyrosynchrotron emission from any of the O-type stars. This is almost certainly because the radio photospheres of the magnetic O-type stars are larger than their magnetospheres, and therefore absorb any radio emission \citep{2015MNRAS.452.1245C}. Detailed individual studies of the extremely rapid rotators HR\,5907 and HR\,7355 were conducted by \cite{2017MNRAS.467.2820L,2018MNRAS.tmp..240L}, who found both stars to be exceptionally bright in the radio. They furthermore offered the tantalizing suggestion that their X-ray emission might contain a non-thermal auroral component in addition to the thermal component from the MCWS, with the aurorae originating from the same high-energy electrons that generate the gyrosynchrotron emission. Extremely hard X-ray emission has also been reported by \cite{2018A&A...619A..33R} for the late Bp star CU\,Vir, who also suggested the possibility of a non-thermal auroral component.

Perhaps the most exciting result in radio has been the growth in the number known `radio lighthouse' stars. Until recently only one such  star was known, namely CU\,Vir, however the last few years have seen four more discoveries (see \citealt{2015MNRAS.452.1245C,2018MNRAS.474L..61D,2019MNRAS.489L.102D,2019ApJ...877..123D,2019MNRAS.482L...4L}; see also Das et al., this volume), indicating that this phenomenon may in fact be quite common. 

Building on the ORM and magnetospheric parameters derived by \cite{2019MNRAS.490..274S}, a detailed investigation of the H$\alpha$ emission properties of the early B-type stars has yielded definitive evidence that the threshold of emission onset, the scaling of emission strength, and the detailed shapes of emission lines can only be explained only via centrifugal breakout. These results will be published by Shultz et al. (in prep.) and Owocki et al. (in prep.), however a preview is given by Owocki et al. (this volume). 

\acknowledgements{MES acknowledges the financial support provided by the Annie Jump Cannon Fellowship, supported by the University of Delaware and endowed by the Mount Cuba Astronomical Observatory.}




\bibliographystyle{ptapap}
\bibliography{shultz}

\begin{thebibliography}{88}
\providecommand{\natexlab}[1]{#1}
\providecommand{\url}[1]{\texttt{#1}}
\providecommand{\urlprefix}{URL }
\providecommand{\eprint}[2][]{\url{#2}}

\bibitem[{{Auri{\`e}re} et~al.(2007)}]{2007A&A...475.1053A}
{Auri{\`e}re}, M., et~al., \emph{\aap} \textbf{475}, 1053 (2007)

\bibitem[{{Babel} \& {Montmerle}(1997)}]{bm1997}
{Babel}, J., {Montmerle}, T., \emph{\apjl} \textbf{485}, L29 (1997)

\bibitem[{{Bagnulo} et~al.(2006)}]{bagn2006}
{Bagnulo}, S., et~al., \emph{\aap} \textbf{450}, 777 (2006)

\bibitem[{{Bard} \& {Townsend}(2016)}]{2016MNRAS.462.3672B}
{Bard}, C., {Townsend}, R. H.~D., \emph{\mnras} \textbf{462}, 4, 3672 (2016)

\bibitem[{{Bohlender} \& {Monin}(2011)}]{bohl2011}
{Bohlender}, D.~A., {Monin}, D., \emph{\aj} \textbf{141}, 169 (2011)

\bibitem[{{Braithwaite} \& {Spruit}(2004)}]{2004Natur.431..819B}
{Braithwaite}, J., {Spruit}, H.~C., \emph{\nat} \textbf{431}, 819 (2004)

\bibitem[{{Castro} et~al.(2017)}]{2017A&A...597L...6C}
{Castro}, N., et~al., \emph{\aap} \textbf{597}, L6 (2017)

\bibitem[{{Chandra} et~al.(2015)}]{2015MNRAS.452.1245C}
{Chandra}, P., et~al., \emph{\mnras} \textbf{452}, 1245 (2015)

\bibitem[{{Daley-Yates} et~al.(2019){Daley-Yates}, {Stevens}, \&
  {ud-Doula}}]{2019MNRAS.489.3251D}
{Daley-Yates}, S., {Stevens}, I.~R., {ud-Doula}, A., \emph{\mnras}
  \textbf{489}, 3, 3251 (2019)

\bibitem[{{Das} et~al.(2019{\natexlab{a}}){Das}, {Chandra}, {Shultz}, \&
  {Wade}}]{2019ApJ...877..123D}
{Das}, B., {Chandra}, P., {Shultz}, M.~E., {Wade}, G.~A., \emph{\apj}
  \textbf{877}, 2, 123 (2019{\natexlab{a}})

\bibitem[{{Das} et~al.(2019{\natexlab{b}}){Das}, {Chandra}, {Shultz}, \&
  {Wade}}]{2019MNRAS.489L.102D}
{Das}, B., {Chandra}, P., {Shultz}, M.~E., {Wade}, G.~A., \emph{\mnras}
  \textbf{489}, 1, L102 (2019{\natexlab{b}})

\bibitem[{{Das} et~al.(2018){Das}, {Chandra}, \& {Wade}}]{2018MNRAS.474L..61D}
{Das}, B., {Chandra}, P., {Wade}, G.~A., \emph{\mnras} \textbf{474}, L61 (2018)

\bibitem[{{David-Uraz} et~al.(2019)}]{2019MNRAS.483.2814D}
{David-Uraz}, A., et~al., \emph{\mnras} \textbf{483}, 2, 2814 (2019)

\bibitem[{{Drake} et~al.(1987)}]{1987ApJ...322..902D}
{Drake}, S.~A., et~al., \emph{\apj} \textbf{322}, 902 (1987)

\bibitem[{{Duez} et~al.(2010){Duez}, {Braithwaite}, \&
  {Mathis}}]{2010ApJ...724L..34D}
{Duez}, V., {Braithwaite}, J., {Mathis}, S., \emph{\apjl} \textbf{724}, L34
  (2010)

\bibitem[{{Eikenberry} et~al.(2014)}]{2014ApJ...784L..30E}
{Eikenberry}, S.~S., et~al., \emph{\apjl} \textbf{784}, L30 (2014)

\bibitem[{{Fossati} et~al.(2016)}]{2016A&A...592A..84F}
{Fossati}, L., et~al., \emph{\aap} \textbf{592}, A84 (2016)

\bibitem[{{Grunhut} et~al.(2012)}]{grun2012}
{Grunhut}, J.~H., et~al., \emph{\mnras} \textbf{419}, 1610 (2012)

\bibitem[{{Grunhut} et~al.(2013)}]{2013MNRAS.428.1686G}
{Grunhut}, J.~H., et~al., \emph{\mnras} \textbf{428}, 1686 (2013)

\bibitem[{{Grunhut} et~al.(2017)}]{2017MNRAS.465.2432G}
{Grunhut}, J.~H., et~al., \emph{\mnras} \textbf{465}, 2432 (2017)

\bibitem[{{Keszthelyi} et~al.(2019)}]{2019MNRAS.485.5843K}
{Keszthelyi}, Z., et~al., \emph{\mnras} \textbf{485}, 4, 5843 (2019)

\bibitem[{{Kochukhov} \& {Bagnulo}(2006)}]{2006AA...450..763K}
{Kochukhov}, O., {Bagnulo}, S., \emph{\aap} \textbf{450}, 763 (2006)

\bibitem[{{Kochukhov} et~al.(2011){Kochukhov}, {Lundin}, {Romanyuk}, \&
  {Kudryavtsev}}]{koch2011}
{Kochukhov}, O., {Lundin}, A., {Romanyuk}, I., {Kudryavtsev}, D., \emph{ApJ}
  \textbf{726}, 24 (2011)

\bibitem[{{Kochukhov} et~al.(2019){Kochukhov}, {Shultz}, \&
  {Neiner}}]{2019A&A...621A..47K}
{Kochukhov}, O., {Shultz}, M., {Neiner}, C., \emph{\aap} \textbf{621}, A47
  (2019)

\bibitem[{{Krti{\v{c}}ka} et~al.(2017)}]{2017MNRAS.464..933K}
{Krti{\v{c}}ka}, J., et~al., \emph{\mnras} \textbf{464}, 1, 933 (2017)

\bibitem[{{Kurapati} et~al.(2017)}]{2017MNRAS.465.2160K}
{Kurapati}, S., et~al., \emph{\mnras} \textbf{465}, 2160 (2017)

\bibitem[{{Landstreet}(1990)}]{1990ApJ...352L...5L}
{Landstreet}, J.~D., \emph{\apjl} \textbf{352}, L5 (1990)

\bibitem[{{Landstreet} \& {Mathys}(2000)}]{landmat2000}
{Landstreet}, J.~D., {Mathys}, G., \emph{\aap} \textbf{359}, 213 (2000)

\bibitem[{{Landstreet} et~al.(2007)}]{land2007}
{Landstreet}, J.~D., et~al., \emph{\aap} \textbf{470}, 685 (2007)

\bibitem[{{Landstreet} et~al.(2008)}]{land2008}
{Landstreet}, J.~D., et~al., \emph{\aap} \textbf{481}, 465 (2008)

\bibitem[{{Leone} et~al.(2010)}]{leone2010}
{Leone}, F., et~al., \emph{\mnras} \textbf{401}, 2739 (2010)

\bibitem[{{Leto} et~al.(2017)}]{2017MNRAS.467.2820L}
{Leto}, P., et~al., \emph{\mnras} \textbf{467}, 2820 (2017)

\bibitem[{{Leto} et~al.(2018)}]{2018MNRAS.tmp..240L}
{Leto}, P., et~al., \emph{\mnras}  (2018)

\bibitem[{{Leto} et~al.(2019)}]{2019MNRAS.482L...4L}
{Leto}, P., et~al., \emph{\mnras} \textbf{482}, 1, L4 (2019)

\bibitem[{{Linsky} et~al.(1992){Linsky}, {Drake}, \&
  {Bastian}}]{1992ApJ...393..341L}
{Linsky}, J.~L., {Drake}, S.~A., {Bastian}, T.~S., \emph{\apj} \textbf{393},
  341 (1992)

\bibitem[{{Mikul{\'a}{\v s}ek} et~al.(2008)}]{miku2008}
{Mikul{\'a}{\v s}ek}, Z., et~al., \emph{\aap} \textbf{485}, 585 (2008)

\bibitem[{{Mikul{\'a}{\v s}ek} et~al.(2011)}]{miku2011}
{Mikul{\'a}{\v s}ek}, Z., et~al., \emph{\aap} \textbf{534}, L5 (2011)

\bibitem[{{Mikul{\'a}{\v s}ek} et~al.(2018)}]{2018CoSka..48..203M}
{Mikul{\'a}{\v s}ek}, Z., et~al., \emph{Contributions of the Astronomical
  Observatory Skalnate Pleso} \textbf{48}, 203 (2018)

\bibitem[{{Mikul{\'a}{\v s}ek} et~al.(2017)}]{2017ASPC..510..220M}
{Mikul{\'a}{\v s}ek}, Z.~Z., et~al., in Y.~Y. {Balega}, D.~O. {Kudryavtsev},
  I.~I. {Romanyuk}, I.~A. {Yakunin} (eds.) Stars: From Collapse to Collapse,
  \emph{Astronomical Society of the Pacific Conference Series}, volume 510, 220
  (2017)

\bibitem[{{Mikul{\'a}{\v{s}}ek} et~al.(2019)}]{2019arXiv191204121M}
{Mikul{\'a}{\v{s}}ek}, Z., et~al., \emph{arXiv e-prints} arXiv:1912.04121
  (2019)

\bibitem[{{Munoz} et~al.(2019)}]{2019MNRAS.tmp.2573M}
{Munoz}, M.~S., et~al., \emph{\mnras} 2573 (2019)

\bibitem[{{Naz{\'e}} et~al.(2014)}]{2014ApJS..215...10N}
{Naz{\'e}}, Y., et~al., \emph{\apjs} \textbf{215}, 10 (2014)

\bibitem[{{Neiner} et~al.(2015)}]{2015IAUS..305...61N}
{Neiner}, C., et~al., in K.~N. {Nagendra}, S.~{Bagnulo}, R.~{Centeno},
  M.~{Jes{\'u}s Mart{\'{\i}}nez Gonz{\'a}lez} (eds.) Polarimetry, \emph{IAU
  Symposium}, volume 305, 61--66 (2015)

\bibitem[{{Oksala} et~al.(2012)}]{oks2012}
{Oksala}, M.~E., et~al., \emph{MNRAS} \textbf{419}, 959 (2012)

\bibitem[{{Oksala} et~al.(2015{\natexlab{a}})}]{2015A&A...578A.112O}
{Oksala}, M.~E., et~al., \emph{\aap} \textbf{578}, A112 (2015{\natexlab{a}})

\bibitem[{{Oksala} et~al.(2015{\natexlab{b}})}]{2015MNRAS.451.2015O}
{Oksala}, M.~E., et~al., \emph{\mnras} \textbf{451}, 2015 (2015{\natexlab{b}})

\bibitem[{{Oskinova} et~al.(2011)}]{oskinova2011}
{Oskinova}, L.~M., et~al., \emph{\mnras} \textbf{416}, 1456 (2011)

\bibitem[{{Oskinova} et~al.(2014)}]{2014NatCo...5E4024O}
{Oskinova}, L.~M., et~al., \emph{Nature} \textbf{5}, 4024 (2014)

\bibitem[{{Owocki} \& {Cranmer}(2018)}]{2018MNRAS.474.3090O}
{Owocki}, S.~P., {Cranmer}, S.~R., \emph{\mnras} \textbf{474}, 3090 (2018)

\bibitem[{{Owocki} et~al.(2016)}]{2016MNRAS.462.3830O}
{Owocki}, S.~P., et~al., \emph{\mnras} \textbf{462}, 3830 (2016)

\bibitem[{{Petit} \& {Wade}(2012)}]{petit2012a}
{Petit}, V., {Wade}, G.~A., \emph{MNRAS} \textbf{420}, 773 (2012)

\bibitem[{{Petit} et~al.(2013)}]{petit2013}
{Petit}, V., et~al., \emph{\mnras} \textbf{429}, 398 (2013)

\bibitem[{{Petit} et~al.(2017)}]{2017MNRAS.466.1052P}
{Petit}, V., et~al., \emph{\mnras} \textbf{466}, 1052 (2017)

\bibitem[{{Petit} et~al.(2019)}]{2019MNRAS.489.5669P}
{Petit}, V., et~al., \emph{\mnras} \textbf{489}, 4, 5669 (2019)

\bibitem[{{Pillitteri} et~al.(2017){Pillitteri}, {Wolk}, {Reale}, \&
  {Oskinova}}]{2017A&A...602A..92P}
{Pillitteri}, I., {Wolk}, S.~J., {Reale}, F., {Oskinova}, L., \emph{\aap}
  \textbf{602}, A92 (2017)

\bibitem[{{Pillitteri} et~al.(2018)}]{2018A&A...610L...3P}
{Pillitteri}, I., et~al., \emph{\aap} \textbf{610}, L3 (2018)

\bibitem[{{Piskunov} \& {Kochukhov}(2002)}]{pk2002}
{Piskunov}, N.~E., {Kochukhov}, O., \emph{\aap} \textbf{381}, 736 (2002)

\bibitem[{{Preston}(1967)}]{preston1967}
{Preston}, G.~W., \emph{\apj} \textbf{150}, 547 (1967)

\bibitem[{{Rivinius} et~al.(2013)}]{rivi2013}
{Rivinius}, T., et~al., \emph{\mnras} \textbf{429}, 177 (2013)

\bibitem[{{Robrade} et~al.(2018)}]{2018A&A...619A..33R}
{Robrade}, J., et~al., \emph{\aap} \textbf{619}, A33 (2018)

\bibitem[{{Shultz} et~al.(2016)}]{2016ASPC..506..305S}
{Shultz}, M., et~al., in T.~A.~A. {Sigut}, C.~E. {Jones} (eds.) Bright
  Emissaries: Be Stars as Messengers of Star-Disk Physics, \emph{Astronomical
  Society of the Pacific Conference Series}, volume 506, 305 (2016)

\bibitem[{{Shultz} et~al.(2017)}]{2017MNRAS.471.2286S}
{Shultz}, M., et~al., \emph{\mnras} \textbf{471}, 2286 (2017)

\bibitem[{{Shultz} et~al.(2018{\natexlab{a}})}]{2018MNRAS.475..839S}
{Shultz}, M., et~al., \emph{\mnras} \textbf{475}, 839 (2018{\natexlab{a}})

\bibitem[{{Shultz} et~al.(2019{\natexlab{a}})}]{2019MNRAS.486.5558S}
{Shultz}, M., et~al., \emph{\mnras} \textbf{486}, 4, 5558 (2019{\natexlab{a}})

\bibitem[{{Shultz} et~al.(2018{\natexlab{b}})}]{2018MNRAS.475.5144S}
{Shultz}, M.~E., et~al., \emph{\mnras} \textbf{475}, 5144 (2018{\natexlab{b}})

\bibitem[{{Shultz} et~al.(2019{\natexlab{b}})}]{2019MNRAS.490..274S}
{Shultz}, M.~E., et~al., \emph{\mnras} \textbf{490}, 1, 274
  (2019{\natexlab{b}})

\bibitem[{{Shultz} et~al.(2019{\natexlab{c}})}]{2019MNRAS.485.1508S}
{Shultz}, M.~E., et~al., \emph{\mnras} \textbf{485}, 1508 (2019{\natexlab{c}})

\bibitem[{{Sikora} et~al.(2019{\natexlab{a}}){Sikora}, {Wade}, {Power}, \&
  {Neiner}}]{2019MNRAS.483.2300S}
{Sikora}, J., {Wade}, G.~A., {Power}, J., {Neiner}, C., \emph{\mnras}
  \textbf{483}, 2300 (2019{\natexlab{a}})

\bibitem[{{Sikora} et~al.(2019{\natexlab{b}}){Sikora}, {Wade}, {Power}, \&
  {Neiner}}]{2019MNRAS.483.3127S}
{Sikora}, J., {Wade}, G.~A., {Power}, J., {Neiner}, C., \emph{\mnras}
  \textbf{483}, 3127 (2019{\natexlab{b}})

\bibitem[{{Sikora} et~al.(2015)}]{2015MNRAS.451.1928S}
{Sikora}, J., et~al., \emph{\mnras} \textbf{451}, 1928 (2015)

\bibitem[{{Sikora} et~al.(2016)}]{2016MNRAS.460.1811S}
{Sikora}, J., et~al., \emph{\mnras} \textbf{460}, 1811 (2016)

\bibitem[{{Stibbs}(1950)}]{1950MNRAS.110..395S}
{Stibbs}, D.~W.~N., \emph{\mnras} \textbf{110}, 395 (1950)

\bibitem[{{Townsend}(2008)}]{town2008}
{Townsend}, R.~H.~D., \emph{\mnras} \textbf{389}, 559 (2008)

\bibitem[{{Townsend} \& {Owocki}(2005)}]{town2005c}
{Townsend}, R.~H.~D., {Owocki}, S.~P., \emph{\mnras} \textbf{357}, 251 (2005)

\bibitem[{{Townsend} et~al.(2007){Townsend}, {Owocki}, \&
  {Ud-Doula}}]{town2007}
{Townsend}, R.~H.~D., {Owocki}, S.~P., {Ud-Doula}, A., \emph{\mnras}
  \textbf{382}, 139 (2007)

\bibitem[{{Townsend} et~al.(2010)}]{town2010}
{Townsend}, R.~H.~D., et~al., \emph{\apjl} \textbf{714}, L318 (2010)

\bibitem[{{Townsend} et~al.(2013)}]{town2013}
{Townsend}, R.~H.~D., et~al., \emph{\apj} \textbf{769}, 33 (2013)

\bibitem[{{Trigilio} et~al.(2000)}]{2000A&A...362..281T}
{Trigilio}, C., et~al., \emph{\aap} \textbf{362}, 281 (2000)

\bibitem[{{Trigilio} et~al.(2004)}]{2004A&A...418..593T}
{Trigilio}, C., et~al., \emph{\aap} \textbf{418}, 593 (2004)

\bibitem[{{ud-Doula} \& {Owocki}(2002)}]{ud2002}
{ud-Doula}, A., {Owocki}, S.~P., \emph{ApJ} \textbf{576}, 413 (2002)

\bibitem[{{ud-Doula} et~al.(2008){ud-Doula}, {Owocki}, \& {Townsend}}]{ud2008}
{ud-Doula}, A., {Owocki}, S.~P., {Townsend}, R.~H.~D., \emph{MNRAS}
  \textbf{385}, 97 (2008)

\bibitem[{{ud-Doula} et~al.(2009){ud-Doula}, {Owocki}, \& {Townsend}}]{ud2009}
{ud-Doula}, A., {Owocki}, S.~P., {Townsend}, R.~H.~D., \emph{MNRAS}
  \textbf{392}, 1022 (2009)

\bibitem[{{ud-Doula} et~al.(2006){ud-Doula}, {Townsend}, \& {Owocki}}]{ud2006}
{ud-Doula}, A., {Townsend}, R.~H.~D., {Owocki}, S.~P., \emph{ApJl}
  \textbf{640}, L191 (2006)

\bibitem[{{ud-Doula} et~al.(2013)}]{ud2013}
{ud-Doula}, A., et~al., \emph{\mnras} \textbf{428}, 2723 (2013)

\bibitem[{{ud-Doula} et~al.(2014)}]{ud2014}
{ud-Doula}, A., et~al., \emph{\mnras} \textbf{441}, 3600 (2014)

\bibitem[{{Wade} et~al.(2017)}]{2017MNRAS.465.2517W}
{Wade}, G.~A., et~al., \emph{\mnras} \textbf{465}, 2517 (2017)

\bibitem[{{Weber} \& {Davis}(1967)}]{wd1967}
{Weber}, E.~J., {Davis}, L., Jr., \emph{\apj} \textbf{148}, 217 (1967)

\bibitem[{{Wisniewski} et~al.(2015)}]{2015ApJ...811L..26W}
{Wisniewski}, J.~P., et~al., \emph{\apjl} \textbf{811}, L26 (2015)

\end{thebibliography}

\end{document}